\documentclass[11pt,twoside]{article}
\usepackage{asp2006}
\usepackage{epsf}
\usepackage{graphics}
\usepackage{lscape}
\begin{document}
\title{Beyond the power spectrum: primordial and secondary non-Gaussianity in the microwave background}
\author{Kendrick M.~Smith}
\affil{Department of Astrophysical Sciences, Princeton University}

\begin{abstract}
Cosmic microwave background observations are most commonly analyzed by estimating the power spectrum.
In the limit where the CMB statistics are perfectly Gaussian, this extracts all the information, but
the CMB also contains detectable non-Gaussian contributions from secondary, and possibly primordial,
sources.
We review possible sources of CMB non-Gaussianity and describe statistical techniques which are
optimized for measuring them, complementing the power spectrum analysis.
The machinery of $N$-point correlation functions provides a unifying framework for optimal estimation
of primordial non-Gaussian signals or gravitational lensing.
We review recent results from applying these estimators to data from the WMAP satellite mission.
\end{abstract}

\section{Introduction}

Observations of the cosmic microwave background have had an enormous impact on
our understanding of cosmology and the level of precision with which cosmological parameters
can be constrained.
Beginning with first detection of anisotropy on large angular scales by COBE \citep{Smoot:1992td},
successive generations of experiments have provided increasingly precise measurements of CMB temperature
flucutations on smaller 
scales \citep[e.g.][]{Hanany:2000qf,Netterfield:2001yq,Hinshaw:2003ex,Kuo:2006ya,Ade:2007ty,Lueker:2009rx,Fowler:2010cy}.
In particular, the WMAP satellite has measured CMB fluctuations on the full sky with high signal-to-noise
for all angular scales larger than $\approx$20 arcmin.

It is convenient to represent the CMB in harmonic space.  If $T({\bf n})$ denotes the value of the temperature
in line-of-sight direction ${\bf n}$, then we expand in spherical harmonics:
\begin{equation}
T({\bf n}) = \sum_{\ell=2}^\infty \sum_{m=-\ell}^\ell a_{\ell m} Y_{\ell m}({\bf n})
\end{equation}
to define the multipoles $a_{\ell m}$.  In this representation, the most general rotationally invariant
two-point correlation function is:
\begin{equation}
\left\langle a_{\ell_1 m_1} a_{\ell_2 m_2} \right\rangle = (-1)^{m_1} C_{\ell_1} \delta_{\ell_1\ell_2} \delta_{m_1,-m_2}
\end{equation}
where this equation defines the power spectrum $C_\ell$.

CMB observations are typically analyzed via the power spectrum: the main result of the analysis is a
measurement of $C_\ell$ and an estimate of the statistical uncertainty, which can be compared with theoretical
predictions for the power spectrum.
This approach is powerful because the CMB is a Gaussian field to a good approximation.
Gaussianity implies that the full probability distribution for the CMB map (i.e.~the multivariate PDF for the variables $a_{\ell m}$) 
is determined by the  two-point correlation function, so that the power spectrum contains all the information.
However, there are weak sources of non-Gaussianity which contain qualitatively new cosmological information,
and require statistical techniques which complement the power spectrum.
In this article, we will review sources of non-Gaussianity either in the early universe during inflation,
or in the late universe via gravitational lensing.
Our emphasis will be on statistical techniques for extracting these non-Gaussian signals, and we will present
results from applying these techniques to WMAP data when possible.

\section{Primordial non-Gaussianity}

\subsection{Three-point signals from inflation}

Consider the simplest model of inflation: single-field slow-roll inflation with standard kinetic term,
\begin{equation}
{\mathcal L} = \frac{M_{\rm pl}^2}{2} R - \frac{1}{2} (\partial^\mu \phi) (\partial_\mu \phi) - V(\phi)
\end{equation}
In such a model, the adiabatic curvature perturbation $\zeta({\bf k})$ generated during inflation is a 
Gaussian field; in particular the three-point correlation function is zero.\footnote{This statement is 
an approximation, but a good one: the three-point function in such models corresponds roughly 
to $f_{NL}^{\rm loc} \approx (5/12) (1-n_s)$, far too small to be detectable 
\citep{Acquaviva:2002ud,Maldacena:2002vr}.}
However, there are inflationary models which can generate a three-point function which is observationally
distinguishable from zero.  In this section we will review the phenomenology of such models.

Let us first make some general mathematical comments.  If we restrict attention to models which satisfy
statistical translation and rotation invariance, then the three-point function of the curvature perturbation
$\zeta$ must be of the form
\begin{equation}
\langle \zeta({\bf k}_1) \zeta({\bf k}_2) \zeta({\bf k}_3) \rangle = F(k_1,k_2,k_3) \, (2\pi)^3 \delta^3({\bf k}_1 + {\bf k}_2 + {\bf k}_3)
\end{equation}
where the function $F$ is called the bispectrum, and only depends on the lengths $k_i$ of the Fourier
wavenumbers ${\bf k}_i$ which form a closed triangle.
For the inflationary models we will consider in this section, the bispectrum will have a simple dependence on
overall scale (roughly $F(\alpha k_1, \alpha k_2, \alpha k_3) \approx \alpha^{-6 + 2(1-n_s)} F(k_1,k_2,k_3)$)
so that we can think of the bispectrum as a function of a 2-parameter family of triangle shapes without regard
to overall scale.
We will use the term ``squeezed'' for a triangle shape with $k_1\ll k_2$ (which implies $k_1\ll k_3$),
``flattened'' for a shape with $k_1 \approx (k_2+k_3)$, and ``equilateral'' for a shape with
$k_1\approx k_2\approx k_3$.

The curvaton model is a multifield model of inflation in which the source of primordial curvature fluctuations
is not the inflaton, but a second field $\sigma$ which does not dominate the energy density during inflation,
but decays after the inflaton, when the curvaton is oscillating near the minimum of its potential
\citep{Linde:1996gt,Moroi:2001ct,Lyth:2002my}.
The power spectrum $P_{(\delta\sigma/\sigma)}$ of the fractional curvaton perturbation $(\delta\sigma/\sigma)$
and the fraction $f = \rho_\sigma/\rho_{\rm tot}$ of the energy density due to the curvaton (both quantities
evaluated at curvaton decay) are free parameters of the model.
The perturbation to the curvaton energy density is given 
by $(\delta\rho_\sigma/\rho_\sigma) = 2(\delta\sigma/\sigma) + (\delta\sigma/\sigma)^2$,
and can therefore be a non-Gaussian field if the quadratic term is non-negligible compared
to the linear term.
There is a region of parameter space (i.e.~$f \ll 1$ with $P_{\delta\sigma/\sigma} = {\mathcal O}(10^{-9} f^{-2})$)
in which the primordial curvature fluctuation (after curvaton decay) is significantly non-Gaussian and has a power
spectrum which is consistent with CMB observations.
The non-Gaussian curvature fluctuation in the curvaton model can be written:
\begin{equation}
\zeta({\bf x}) = \zeta_G({\bf x}) - \frac{3}{5} f_{NL}^{\rm local} \zeta_G({\bf x})^2
\end{equation}
where $\zeta_G$ is a Gaussian field, and $f_{NL}^{\rm local}$ is a free parameter.
(The factor $3/5$ is purely conventional.)
If we write this equation in Fourier space\footnote{Note that in Eqs.~(\ref{eq:F_local}) and~(\ref{eq:F_equil_orthog}), we have
assumed scale invariance, so that $P_\zeta(k) = \Delta_\zeta k^{-3}$.  See \citet{Senatore:2009gt} for expressions for the primordial
bispectra with a power law spectrum $P_\zeta(k) = \Delta_\zeta k^{-3} (k/k_0)^{n_s-1}$.}, we get the following form of the bispectrum $F$:
\begin{equation}
F(k_1,k_2,k_3) = f_{NL}^{\rm local} F_{\rm local}(k_1,k_2,k_3)
\end{equation}
where
\begin{equation}
F_{\rm local}(k_1,k_2,k_3) = -\frac{6 \Delta_\zeta^2}{5} \left( 
  \frac{1}{k_1^3 k_2^3} +
  \frac{1}{k_2^3 k_3^3} +
  \frac{1}{k_1^3 k_3^3} \right)  \label{eq:F_local}
\end{equation}
The local bispectrum is largest for squeezed triangles.
Conversely, there is a theorem, the single-field consistency relation \citep{Maldacena:2002vr,Creminelli:2004yq},
which states that the bispectrum is always small (i.e.~${\mathcal O}(10^{-2})$)
in squeezed triangles, provided that inflation is single-field (but allowing
for aribtrary self-interactions of the inflaton).
Thus the presence of a second field in the curvaton model is actually a necessary ingredient for
any model which generates $f_{NL}^{\rm local}$ larger than ${\mathcal O}(10^{-2})$.

For single-field inflation, there is also a theorem \citep{Senatore:2009gt}
which shows that the most general primordial bispectrum is:
\begin{equation}
F(k_1,k_2,k_3) = f_{NL}^{\rm equil} F_{\rm equil}(k_1,k_2,k_3) + f_{NL}^{\rm orthog} F_{\rm orthog}(k_1,k_2,k_3)  \label{eq:F_single_field}
\end{equation}
where the ``equilateral'' and ``orthogonal'' bispectra are defined by:\footnote{The equilateral and orthogonal
bispectra defined in Eq.~(\ref{eq:F_equil_orthog}) are actually approximations to more precise expressions for the bispectra
generated during inflation.  The approximation is made so that $F(k_1,k_2,k_3)$ will be a sum of a small number
of terms which are factorizable in the form $f(k_1) g(k_2) h(k_3)$.  As will be discussed in the next subsection,
this factorizability condition is necessary in order to make the data analysis computationally tractable.}
\begin{eqnarray}
F_{\rm equil}(k_1,k_2,k_3) &=& - \frac{18 \Delta_\zeta^2}{5} \frac{(k_1+k_2-k_3)(k_1-k_2+k_3)(-k_1+k_2+k_3)}{k_1^3 k_2^3 k_3^3} \nonumber \\
F_{\rm orthog}(k_1,k_2,k_3) &=& 3 F_{\rm equil}(k_1,k_2,k_3) + \frac{36 \Delta_\zeta^2}{5} \frac{1}{k_1^2 k_2^2 k_3^2}  \label{eq:F_equil_orthog}
\end{eqnarray}
The equilateral shape is largest for equilateral triangles (as the name suggests) while the orthogonal shape changes sign between equilateral
and flattened triangles.  Both single-field shapes vanish in the squeezed limit, as required by the consistency relation.

For a particular single-field model, such as DBI inflation \citep{Alishahiha:2004eh}
or ghost inflation \citep{ArkaniHamed:2003uz}, the coefficients $f_{NL}^{\rm equil}$ and
$f_{NL}^{\rm orthog}$ in Eq.~(\ref{eq:F_single_field}) can be calculated in terms
of fundamental parameters of the model.
In this article, our emphasis will be on data analysis.
We simply remark that the forms of the bispectra $F_{\rm local}$, $F_{\rm equil}$, $F_{\rm orthog}$
given above provide a point of contact between theory and data.
Given data from a CMB experiment like WMAP, our job is to determine observational
limits on the three $f_{NL}$ parameters with statistical errors which are as small
as possible.
Once this has been done, any inflationary model can be compared with observations by
calculating the $f_{NL}$ parameters.

Primordial non-Gaussianity is a particularly interesting probe of inflation because it can
rule out qualitative classes of models.
A robust detection of $f_{NL}^{\rm local} \ne 0$ would rule out all single-field
models.
Detection of any nonzero $f_{NL}$ parameter would rule out the simplest inflationary model,
single-field inflation with standard kinetic term and slow-roll potential $V(\phi)$.
In the next two subsections we will turn to data analysis: we will construct optimal estimators
for constraining the $f_{NL}$ parameters from data, and present results from WMAP.

\subsection{Estimators for primordial non-Gaussianity}
\label{ssec:fnl_estimators}

Let us begin by making some general statements about how three-point signals are estimated from data,
before specializing to the case of the three inflationary shapes from the preceding subsection.

We will make the approximation that the evolution
from the 3D intitial curvature fluctuation $\zeta({\bf k})$ to the 2D CMB $a_{\ell m}$
is a linear operation, so that the three-point function of $\zeta$ translates linearly
into the three-point function of the CMB:
\begin{equation}
\left\langle a_{\ell_1 m_1} a_{\ell_2 m_2} a_{\ell_3 m_3} \right\rangle
  = f_{NL}^X B^X_{\ell_1m_1,\ell_2m_2,\ell_3m_3}  \label{eq:fnlhat_setup}
\end{equation}
where $B^X_{\ell_1m_1,\ell_2m_2,\ell_3m_3}$ is different for each $X\in\{ {\rm local}, {\rm equil}, {\rm orthog} \}$.

When we measure the CMB in a real experiment, we measure the $a_{\ell m}$'s plus some instrumental noise:
\begin{equation}
a_{\ell m}^{\rm obs} = a_{\ell m}^{\rm true} + a_{\ell m}^{\rm noise}
\end{equation}
We will assume that the instrumental noise is Gaussian, and that the total covariance matrix of the observed $a_{\ell m}$'s
is given by the sum of a signal term (which is diagonal in $\ell,m$) and a noise term (which is non-diagonal):
\begin{equation}
\left\langle a_{\ell_1m_1} a_{\ell_2m_2} \right\rangle = (-1)^{m_1} C_{\ell_1} \delta_{\ell_1\ell_2} \delta_{m_1,-m_2} + N_{\ell_1 m_1,\ell_2 m_2} \label{eq:ctot}
\end{equation}
The detailed form of the noise covariance matrix $N_{\ell_1m_1,\ell_2m_2}$ encapsulates details of the experiment such as
spatial gradients in instrumental noise level (represented by assigning different noise variance to different pixels), the
sky mask imposed to remove regions of high foreground emission (represented by assigning infinite noise variance to pixels
which are masked), and the shapes of the beams.
We will denote the total covariance matrix on the RHS of Eq.~(\ref{eq:ctot}) by $C_{\ell_1m_1,\ell_2m_2}$ and its inverse by $(C^{-1})^{\ell_1m_1,\ell_2m_2}$.
The matrix $C$ represents all the properties of the experiment that we will need to know about in order to write down
an estimator for $f_{NL}$.

Now we can ask the following general question:
Given $B^X_{\ell_1m_1,\ell_2m_2,\ell_3m_3}$ and $C_{\ell_1m_1,\ell_2m_2}$, how should we estimate the value
of the coefficient $f_{NL}^X$ in Eq.~(\ref{eq:fnlhat_setup}) from the noisy CMB $a_{\ell m}$?
The optimal (i.e. mininum variance) estimator was found in \citep{Creminelli:2005hu}:
\begin{equation}
{\hat f_{NL}}^X = \frac{1}{F} \left( T[C^{-1}a] - (C^{-1}a)^{\ell m} 
   \left\langle \partial_{\ell m} T[C^{-1}a'] \right\rangle_{a'} \right)  \label{eq:fnlhat}
\end{equation}
where the quantities $T$ and $\partial T$ are defined by \citep{Smith:2006ud}:
\begin{eqnarray}
T[C^{-1}a] &=& \frac{1}{6} \sum_{\ell_im_i} B_{\ell_1m_1,\ell_2m_2,\ell_3m_3} (C^{-1}a)^{\ell_1m_1} (C^{-1}a)^{\ell_2m_2} (C^{-1}a)^{\ell_3m_3}  \nonumber \\
\partial_{\ell m} T[C^{-1}a] &=& \frac{1}{2} \sum_{\ell_im_i} B_{\ell m,\ell_1m_1,\ell_2m_2} (C^{-1}a)^{\ell_1m_1} (C^{-1}a)^{\ell_2m_2}  \label{eq:Tdef}
\end{eqnarray}
The normalization constant $F$, which is included so that ${\hat f_{NL}}^X$ will be an unbiased estimator of $f_{NL}^X$ (i.e.
$\langle {\hat f_{NL}}^X \rangle = f_{NL}^X$) can be determined by Monte Carlo.
The estimator ${\hat f_{NL}}^X$ in Eq.~(\ref{eq:fnlhat}) is a sum of a three-point term and a one-point ``counterterm'' which
reduces the estimator variance in the presence of inhomogeneous noise.
The one-point term vanishes if the noise is homogeneous, since rotation invariance implies $\langle \partial_{\ell m} T[C^{-1}a'] \rangle = 0$
for $\ell > 0$.
We will see more examples of this estimator structure, with an $N$-point leading term plus counterterms of lower order,
in \S\ref{ssec:gravitational_lensing_estimators}

To evaluate the estimator ${\hat f_{NL}}^X$, we need algorithms for computing $a_{\ell m} \rightarrow (C^{-1}a)^{\ell m}$
and $a \rightarrow T[a]$.
The first algorithm depends on details of the experiment being analyzed (via $C^{-1}$) but not the
form of the three-point function  $B_{\ell_1m_1,\ell_2m_2,\ell_3m_3}$; we will discuss its implementation in
WMAP in the next subsection.
The rest of this subsection is devoted to briefly describing the second algorithm, which depends only on the three-point function.

For a completely general three-point function $B_{\ell_1m_1,\ell_2m_2,\ell_3m_3}$, there is no algorithm for the
operation $a \rightarrow T[a]$ which is faster than the ${\mathcal O}(\ell_{\rm max}^5)$ harmonic-space sum
in Eq.~(\ref{eq:Tdef}).  For current CMB experiments (with $\ell_{\rm max} = {\mathcal O}(10^3)$) this is computationally
prohibitive.  However, if the three-point function is a sum of $N$ factorizable terms, in the sense that
\begin{eqnarray}
B_{\ell_1m_1,\ell_2m_2,\ell_3m_3} &=& \sum_{i=1}^N 
  \alpha_{\ell_1\ell_2\ell_3}^{(i)}
  \beta_{\ell_1\ell_2\ell_3}^{(i)}
  \gamma_{\ell_1\ell_2\ell_3}^{(i)} \label{eq:B_factorizable} \\
&& \hspace{0.5cm} \times
  \sqrt{\frac{(2\ell_1+1)(2\ell_2+1)(2\ell_3+1)}{4\pi}}
  \left( \begin{array}{ccc} 
     \ell_1 & \ell_2 & \ell_3 \\
        m_1 &   m_2  & m_3
  \end{array} \right)  \nonumber 
\end{eqnarray}
and $N$ is not too large, then there is a fast (i.e.~${\mathcal O}(\ell_{\rm max}^3 N)$) algorithm
for the $a\rightarrow T[a]$ operation \citep{Komatsu:2003iq,Creminelli:2005hu,Smith:2006ud}.
(A closely related algorithm exists for the operation $a\rightarrow \partial_{\ell m}T[a]$, which
is also needed in order to evaluate the one-point term in the optimal estimator in Eq.~(\ref{eq:fnlhat}).)
In practice, this factorizability condition is satisfied for ``interesting'' forms of the three-point function such as the 
local, equilateral and orthogonal shapes.

In a little more detail, we can also define a 3D factorizability condition for the primordial bispectrum
$F(k_1,k_2,k_3)$: we say that $F$ is factorizable if it is a sum of $N$ terms
\begin{equation}
F(k_1,k_2,k_3) = \sum_{i=1}^N \alpha_i(k_1) \beta_i(k_2) \gamma_i(k_3)   \label{eq:F_factorizable}
\end{equation}
It can be shown
that a primoridal bispectrum $F$ which is factorizable (in the sense defined by Eq.~(\ref{eq:F_factorizable}))
evolves to a CMB three-point function which is also factorizable (in the sense defined by Eq.~(\ref{eq:B_factorizable})),
although the number of terms $N$ is typically increased by a factor 10--100.\footnote{More precisely,
each factorizable term in the primordial bispectrum evolves to a CMB three-point function which has an
integral representation with factorizable integrad.  When the integral is approximated by a finite sum,
the CMB three-point function will be factorizable in the sense defined by Eq.~(\ref{eq:B_factorizable}).
There is an optimization algorithm \citep{Smith:2006ud} which can be used to minimize the number of quadrature
points needed to approximate the integral to a specified level of accuracy.}

The optimal estimator can also be used to search for three-point signals which are generated
by secondary sources of anisotropy such as point sources or gravitational lensing (e.g.
\citet{Spergel:1999xn,Goldberg:1999xm,Verde:2002mu,Cooray:1999kg}).
In this case there is no underlying primordial bispectrum $F$, but the CMB three-point function
$B_{\ell_1m_1,\ell_2m_2,\ell_3m_3}$ seems to satisfy the factorizability condition for a wide
range of secondaries (for a survey of shapes see \citet{Smith:2006ud}).

Summarizing, there is an optimal estimator ${\widehat \mathcal E}$ which can be applied to estimate
the amplitude of a three-point signal of specified shape.
For the estimator to be computationally feasible, the three-point function must satisfy the technical 
requirement of factorizability (Eq.~(\ref{eq:B_factorizable})).
The factorizability requirement is satisfied for the three primordial shapes we will consider in this
article (and for many secondary shapes as well).
Evaluating the estimator also requires a fast algorithm for the experiment-specific operation 
$a_{\ell m} \rightarrow (C^{-1}a)^{\ell m}$ which we now discuss in more detail in the context
of WMAP.

\subsection{Primordial non-Gaussianity in WMAP}
\label{ssec:fnl_wmap}

In the previous subsection, the computational problem of evaluating the optimal estimator ${\hat f_{NL}}^X$ was
reduced to the ``$C^{-1}$ problem'': finding an efficient algorithm to compute the inverse (signal+noise) weighted
map $(C^{-1}a)^{\ell m}$ given a harmonic-space map $a_{\ell m}$.

The operator $C^{-1}$ is experiment-specific.
One simplifying feature of WMAP is that the instrumental noise can be treated as uncorrelated between pixels
(more precisely, pixel correlations due to $1/f$ noise in the detectors are only important on large angular
scales, where the instrumental noise is much smaller than the CMB temperature fluctuations).
Even with this simplification, the $C^{-1}$ problem is still very challenging for WMAP, due to the large number
of pixels ($N_{\rm pix} \approx 10^7$) needed to represent the WMAP data.
A brute-force linear algebra approach, representing $C^{-1}$ by a dense $N_{\rm pix}$-by-$N_{\rm pix}$ matrix, 
is computationally infeasible.
Consequently, algorithms for solving the $C^{-1}$ problem for a large-$N_{\rm pix}$ experiment such as WMAP
are based on iterative methods such as the conjugate gradient algorithm.

Iterative algorithms can be used if there is a fast algorithm for applying the ``forward'' operation
$a \rightarrow Ca$, and a fast algorithm for computing a linear operation $a \rightarrow Pa$ (the
``preconditioner'') such that $P$ approximates $C^{-1}$ as closely as possible.
The idea behind the iterative algorithms is to compute $C^{-1}a$ by iteratively improving a guess $x \approx C^{-1}a$.
In each iteration, the preconditioner is applied to the residual vector $(a - Cx)$ to determine a direction
in which to search for an improved guess.
The algorithm terminates when the residual vector is sufficiently close to zero.
The convergence rate is determined by the accuracy of the approximation $C^{-1} \approx P$; thus the
iterative approach will be successful (i.e., fast) if a good preconditioner can be constructed.

The $C^{-1}$ problem is an ingredient in many flavors of optimal estimators, such as optimal power spectrum
estimators, the optimal estimators for primordial non-Gaussianity in this section, and optimal estimators
for gravitational lensing to be considered shortly.
For this reason, the problem of finding a good preconditioner for WMAP has been considered by several authors
\citep[e.g.][]{Oh:1998sr,Eriksen:2004ss,Smith:2007rg}.
The best preconditioner to date was constructed in \citet{Smith:2007rg}, using a multigrid approach: the 
preconditioner for the $C^{-1}$ operation is defined by evaluating a (partially converged) conjugate
gradient $C^{-1}$ operation on a lower-resolution version of the WMAP dataset.
The preconditioner for the lower-resolution $C^{-1}$ operation is defined using even lower resolution,
and so on recursively.
Using the multigrid preconditioner, the operation $a \rightarrow C^{-1} a$ for WMAP can be computed
in $\approx$15 CPU-minutes.
This is sufficiently fast that the optimal estimator ${\hat f_{NL}}^X$ can be evaluated in a Monte Carlo pipeline.

Optimal $f_{NL}$ constraints from WMAP were first reported for the public 5-year release \citep{Smith:2009jr,Senatore:2009gt}.
This implementation of the optimal estimator was subsequently incorporated 
into the WMAP 7-year analysis pipeline \citep{Komatsu:2010fb}.
At the time of this writing, current WMAP constraints on $f_{NL}$ parameters are:
\begin{eqnarray}
f_{NL}^{\rm local} &=& 32 \pm 21 \nonumber \\ 
f_{NL}^{\rm equil} &=& 26 \pm 140 \nonumber \\
f_{NL}^{\rm orthog} &=& -202 \pm 104 \hspace{1cm} \mbox{($1\sigma$ error)} \label{eq:fnl_wmap}
\end{eqnarray}
Thus the WMAP data are consistent with Gaussian initial conditions.
These constraints will improve by a factor $\approx 5$ in a few years with results from
the Planck satellite mission.

\section{Gravitational lensing}
\label{sec:gravitational_lensing}

\subsection{Introduction}

One of the largest sources of ``secondary'' CMB anisotropy, or additional anisotropy generated after recombination,
is gravitational lensing.
For purposes of this article, the effect of gravitational lensing can be succinctly described as follows.
(For more detailed reviews, see \citet{Lewis:2006fu,Hanson:2009kr}.)
Gravitational potentials in the late universe deflect photons, so that an observer who looks in line-of-sight direction ${\bf n}$
sees the part of the surface of last scattering which lies in direction $({\bf n} + \nabla\phi({\bf n}))$, where
the lens potential $\phi({\bf n})$ is a 2D field which can be written as a line-of-sight integral.
Gravitational lensing preserves surface brightness, so that the effect of gravitational lensing is simply to
move CMB aniostropy around,
\begin{equation}
(\Delta \tilde T)({\bf n}) = (\Delta T)({\bf n} + \nabla \phi({\bf n}))
\end{equation}
In this equation and throughout this article, we denote the lensed CMB temperature by $\tilde T({\bf n})$ or ${\tilde a}_{\ell m}$,
and the unlensed temperature by $T({\bf n})$ or $a_{\ell m}$.

Via gravitational lensing, the CMB is indirectly sensitive to the power spectrum $C_\ell^{\phi\phi}$ of
the lens potential.  This power spectrum can be written as a line-of-sight integral which contains geometric
distance factors and the power spectrum $P_\Psi(k)$ of the large-scale Newtonian potential:
\begin{equation}
C_\ell^{\phi\phi} = \frac{8\pi^2}{\ell^3} \int d\chi\, \chi \left( \frac{\chi_*-\chi}{\chi_*\chi} \right)^2 P_\Psi(\chi; k=\ell/\chi)
\end{equation}
In this way, the CMB becomes sensitive to the expansion history and growth of structure in the late universe,
which adds qualitatively new information.  In contrast, the unlensed CMB is very sensitive to a single 
``late universe'' paramter, the angular diameter distance $D_A(z_*)$ to the redshift of recombination, but is otherwise
insensitive to the late universe: this is the so-called angular diameter distance degeneracy \citep{Zaldarriaga:1997ch}.
CMB lensing breaks the angular diameter distance degeneracy and can ultimately provide interesting constraints
on new parameters such as neutrino mass or the dark energy equation of state \citep{Stompor:1998zj,Smith:2006ud}.

How can we reconstruct the lensing power spectrum $C_\ell^{\phi\phi}$ from observations of the CMB?
One possibility is to consider the effect of gravitational lensing on the CMB temperature power spectrum.
Lensing alters $C_\ell^{TT}$ by smoothing the acoustic peaks and adding power in the high-$\ell$ 
tail \citep{Seljak:1995,Hu:2000,Challinor:2005jy}
The peak-smoothing effect can be understood as degree-scale lenses transferring CMB power
between different values of $\ell$, with characteristic scale $\Delta\ell \sim 100$.
Extra power in the damping tail can be understood as small-scale lenses generating
new anisotropy from a smoothly varying CMB backlight.
(On small angular scales, the unlensed CMB power spectrum is exponentially suppressed but
the power spectrum of the lens potential is not).

A more qualitative effect of CMB lensing is that it generates non-Gaussianity.
In the next subsection, we will calculate some higher-point correlation functions due to lensing.

\subsection{Three-point and four-point signals from lensing}

Let us first compute the three-point function $\langle \tilde T \tilde T \phi \rangle$ between the lens
potential and two powers of the (lensed) CMB temperature.
If we Taylor expand the lensed CMB in powers of the lens potential, we get
\begin{equation}
{\tilde a}_{\ell_1 m_1} = a_{\ell_1 m_1} + \sum_{\ell_2m_2\ell m} f_{\ell_1\ell_2\ell}\, a_{\ell_2m_2}^* \phi_{\ell m}^*
\left(
  \begin{array}{ccc}
    \ell_1 & \ell_2 & \ell \\
      m_1 & m_2 & m
  \end{array}
\right) + \cdots
\end{equation}
where the mode-coupling kernel $f_{\ell_1\ell_2\ell}$ is defined by
\begin{eqnarray}
f_{\ell_1\ell_2\ell} &=& 
   \left( \frac{-\ell_1(\ell_1+1) + \ell_2(\ell_2+1) + \ell(\ell+1)}{2} \right)  \nonumber \\
   && \hspace{1cm} \times 
   \sqrt{ \frac{(2\ell_1+1)(2\ell_2+1)(2\ell+1)}{4\pi} }
   \left( \begin{array}{ccc}
     \ell_1 & \ell_2 & \ell \\
        0 & 0 & 0
    \end{array} \right)
\end{eqnarray}
A short calculation now gives the $\langle \tilde T \tilde T \phi \rangle$ three-point function:
\begin{equation}
\left\langle {\tilde a}_{\ell_1 m_1} {\tilde a}_{\ell_2 m_2} \phi_{\ell m} \right\rangle
  = \left( f_{\ell_1\ell_2\ell} C_{\ell_2}^{TT} + f_{\ell_2\ell_1\ell} C_{\ell_1}^{TT} \right) C_\ell^{\phi\phi}
    \left( \begin{array}{ccc}
      \ell_1 & \ell_2 & \ell \\
        m_1 & m_2 & m
     \end{array} \right) \label{eq:lensing_ttphi}
\end{equation}
If we replace the lens potential $\phi_{\ell m}$ with a different large-scale structure tracer field $g_{\ell m}$,
then the correlation between $\phi$ and $g$ will generate a $\langle \tilde T \tilde T g \rangle$ correlation given by:
\begin{equation}
\left\langle {\tilde a}_{\ell_1 m_1} {\tilde a}_{\ell_2 m_2} a_{\ell m}^\phi \right\rangle
  = \left( f_{\ell_1\ell_2\ell} C_{\ell_2}^{TT} + f_{\ell_2\ell_1\ell} C_{\ell_1}^{TT} \right) C_\ell^{g\phi}
    \left( \begin{array}{ccc}
      \ell_1 & \ell_2 & \ell \\
        m_1 & m_2 & m
     \end{array} \right)  \label{eq:lensing_ttg}
\end{equation}
We will consider a case where the tracer field is given by number counts of radio galaxies in the NVSS survey,
in \S\ref{ssec:wmap_lensing} 

We can also consider the case where the tracer field is the CMB temperature itself, 
where the correlation $C_\ell^{T\phi}$ arises from the ISW effect in a cosmology with $\Omega_m \ne 1$.
This gives rise to the ``ISW-lensing'' three-point function, which is internal to the CMB:
\begin{equation}
\left\langle {\tilde a}_{\ell_1m_1} {\tilde a}_{\ell_2m_2} {\tilde a}_{\ell_3m_3} \right\rangle 
  = f_{\ell_1\ell_2\ell_3} C_{\ell_2}^{TT} C_{\ell_3}^{T\phi} 
    \left( \begin{array}{ccc}
      \ell_1 & \ell_2 & \ell_3 \\
        m_1 & m_2 & m_3
     \end{array} \right)
      + \mbox{(5 perm.)}  \label{eq:isw_lensing_ttt}
\end{equation}
The ISW-lensing signal is the main contribution to the three-point function induced by gravitational lensing.
There is a larger signal in the four-point function.
\begin{eqnarray}
&& \left\langle {\tilde a}_{\ell_1 m_1} {\tilde a}_{\ell_2 m_2} {\tilde a}_{\ell_3 m_3} {\tilde a}_{\ell_4 m_4} \right\rangle_{\rm conn} \label{eq:lensing_tttt} \\
&& \hspace{0.5cm} = \sum_{\ell} f_{\ell_1\ell_3\ell} f_{\ell_2\ell_4\ell} C_{\ell_3}^{TT} C_{\ell_4}^{TT} C_\ell^{\phi\phi} \nonumber \\
&& \hspace{1.5cm} \times \sum_m (-1)^m 
    \left( \begin{array}{ccc}
      \ell_1 & \ell_3 & \ell \\
        m_1 & m_3 & m
     \end{array} \right)
    \left( \begin{array}{ccc}
      \ell_2 & \ell_4 & \ell \\
        m_2 & m_4 & -m
     \end{array} \right)
  + \mbox{(11 perm.)} \nonumber
\end{eqnarray}
For comparison, forecasts for the Planck satellite show that the ISW-lensing three-point signal should be detectable at the
$\sim 5\sigma$ level, whereas the four-point signal should be detectable at $\sim 60 \sigma$, and can in fact be used to
make a precision measurement of $C_\ell^{\phi\phi}$.

The three-point and four-point calculations in this section have been somewhat formal, in the sense that we have written down
expressions for the correlation functions (Eqs.~(\ref{eq:lensing_ttphi})--(\ref{eq:lensing_tttt}))
without much interpretation.  In the next section we will give a more intuitive interpretation for these signals,
and also construct estimators for extracting them from data.

\subsection{Estimators for gravitational lensing}
\label{ssec:gravitational_lensing_estimators}

The large higher-point signals described in the preceding subsection contain information, via the lens potential $\phi$,
about the expansion history and growth of structure in the late universe.
How can we best extract this information, i.e. what higher-point estimators should we apply to data to measure these signals?
We will answer this question for the $\langle TTg \rangle$ three-point signal (Eq.~(\ref{eq:lensing_ttg})), the ISW-lensing three-point
signal (Eq.~(\ref{eq:isw_lensing_ttt})), and the four-point CMB signal (Eq.~(\ref{eq:lensing_tttt})).

In order to make the analogy with the estimators for primordial non-Gaussianity the clearest,
let us first consider the ISW-lensing signal.
Since this signal is just a different shape for the CMB three-point function, the minimum-variance estimator ${\widehat \mathcal E}$
is given by the general form in Eq.~(\ref{eq:fnlhat}) given previously in the context of primordial non-Gaussianity.
If we write out the definitions of $T$ and $\partial T$ for the special case of the ISW-lensing three-point function, the
estimator becomes:
\begin{eqnarray}
\widehat{\mathcal E} &=& \frac{1}{F}
  \sum_{\ell_im_i} f_{\ell_1\ell_2\ell_3} C_{\ell_2}^{TT} C_{\ell_3m_3}^{T\phi} 
\left( \begin{array}{ccc}
\ell_1 & \ell_2 & \ell_3 \\
m_1 & m_2 & m_3
\end{array} \right) \nonumber \\
&& \hspace{0.5cm} \times
\Bigg[ (C^{-1}a)^{\ell_1m_1} (C^{-1}a)^{\ell_2m_2} (C^{-1}a)^{\ell_3m_3} \nonumber \\
  && \hspace{1cm} - (C^{-1})^{\ell_1m_1,\ell_2m_2} (C^{-1}a)^{\ell_3m_3} - (C^{-1})^{\ell_1m_1,\ell_3m_3} (C^{-1}a)^{\ell_2m_2} \nonumber \\
  && \hspace{1cm} - (C^{-1})^{\ell_2m_2,\ell_3m_3} (C^{-1}a)^{\ell_1m_1} \Bigg]
\end{eqnarray}
The last two terms in the estimator turn out to be small compared to the first two terms, so we will neglect them.
Under this approximation, we can rewrite $\widehat{\mathcal E}$ in a mathematically equivalent way by introducing 
the quadratic estimator $({\mathcal N}^{-1}{\hat\phi})$, defined\footnote{We have used the notation ${\mathcal N}^{-1}{\hat\phi}$ rather than ${\hat\phi}$ 
because there is a formal sense in which the right-hand side of Eq.~(\ref{eq:phihat})
is a reconstruction for the lens potential after multiplying by the inverse noise covariance matrix of the reconstruction.  More
precisely, there is an operator ${\mathcal N}^{-1}_{\ell m,\ell'm'}$ such that 
$\langle ({\mathcal N}^{-1}{\hat\phi})_{\ell m} \rangle = {\mathcal N}^{-1}_{\ell m,\ell'm'} \phi_{\ell'm'}$
and $\langle ({\mathcal N}^{-1}{\hat\phi})_{\ell m} ({\mathcal N}^{-1}{\hat\phi})_{\ell'm'} \rangle = {\mathcal N}_{\ell m,\ell'm'}$.
In the simplified case of all sky coverage and isotropic noise, the operator ${\mathcal N}^{-1}$ can be inverted and one
can write down an estimator ${\hat\phi}_{\ell m}$ such that $\langle {\hat\phi}_{\ell m} \rangle = \phi_{\ell m}$.
In the cut sky case the operator ${\mathcal N}^{-1}$ is noninvertible; for this reason we work with the inverse noise
weighted reconstruction $({\mathcal N}^{-1}{\hat\phi})_{\ell m}$ in Eq.~(\ref{eq:phihat}) and throughout this article.} by:
\begin{eqnarray}
({\mathcal N}^{-1}{\hat\phi})_{\ell m} &=& (-1)^m \sum_{\ell_1\ell_2\ell} f_{\ell_1\ell_2\ell} C_{\ell_2}^{TT} 
    \left( \begin{array}{ccc}
      \ell_1 & \ell_2 & \ell \\
        m_1 & m_2 & -m
     \end{array} \right) \nonumber \\
&& \times \left[ (C^{-1}a)^{\ell_1m_1} (C^{-1}a)^{\ell_2m_2} - (C^{-1})^{\ell_1m_1,\ell_2m_2} \right]  \label{eq:phihat}
\end{eqnarray}
We have arrived at this definition by studying $N$-point correlation functions, but the quadratic reconstruction ${\mathcal N}^{-1} {\hat\phi}$
was originally proposed as a minimum variance estimator of the lensing potential from the CMB
\citep{Bernardeau:1996aa,Zaldarriaga:1998te,Hu:2001tn}.
Because the lens breaks statistical isotropy of the CMB, correlations between different Fourier modes of the CMB can be used to estimate
the lensing potential.
Each Fourier mode $\phi_{\ell m}$ of the lensing potential can be independently estimated, so the lens reconstruction ${\hat\phi}_{\ell m}$ has
the degrees of freedom of a map, rather than a scalar quantity.
Note that our definition of the estimator in Eq.~(\ref{eq:phihat}) includes subtraction of the ``mean field'' term $(C^{-1})^{\ell_1m_1,\ell_2m_2}$.
This can be interpreted as simply subtracting off the contribution from spurious anisotropy, induced by anisotropy of the noise covariance,
when estimating the lensing potential.

The ISW-lensing estimator ${\mathcal E}$ can then be rewritten:
\begin{equation}
\widehat{\mathcal E} = \frac{1}{F} \sum_{\ell m} C_\ell^{T\phi} ({\mathcal N}^{-1}{\hat\phi})_{\ell m} (C^{-1}a)_{\ell m}^*  \label{eq:isw_lensing_estimator2}
\end{equation}
In this form, the estimator has a simple intuitive interpretation.
The ISW-lensing signal can be interpreted as a correlation between the (inverse noise weighted)
lens reconstruction $({\mathcal N}^{-1}{\hat\phi})$ and the (inverse signal+noise weighted) temperature $(C^{-1}a)$.
The optimal estimator is simply the cross power spectrum of these two fields, evaluated in a single large
bandpower whose ``shape'' in $\ell$ is given by the cross spectrum $C_\ell^{T\phi}$.
When written in this way (Eq.~(\ref{eq:isw_lensing_estimator2})), the estimator looks like a two-point quantity, but it
is actually a three-point statistic in the CMB temperature map,
because the lens reconstruction $({\mathcal N}^{-1}{\hat\phi})$ is quadratic in the CMB.
(Note that there is also a one-point counterterm in the estimator, coming from the mean field term in the
definition of ${\mathcal N}^{-1} {\hat \phi}$.)

The other higher-point signals from the previous subsection can be treated similarly to the ISW-lensing case.
Consider next the $\langle TTg \rangle$ three-point signal (Eq.~(\ref{eq:lensing_ttg})).
In this case, the optimal estimator is given by:
\begin{equation}
\widehat{\mathcal E} = \frac{1}{F} \sum_{\ell m} C_\ell^{\phi g} ({\mathcal N}^{-1}{\hat\phi})_{\ell m} (C^{-1}g)_{\ell m}^*  \label{eq:ttg_estimator}
\end{equation}
For this estimator to make sense, we must have noisy observations of the large-scale structure field 
$g_{\ell m}$, with associated noise and signal covariance so that $(C^{-1}g)_{\ell m}$ can be defined.
Note that the above equation, we have written down the optimal estimator for an overall multiple of
the fiducial signal.
This estimator is just the cross power spectrum of the lens reconstruction ${\hat\phi}_{\ell m}$ and the
observed galaxy field, evaluated in a large bandpower with $\ell$ weighting given by $C_\ell^{\phi g}$.
This is the appropriate estimator for making a statistical detection of a weak signal, but in a case
where the $\langle TTg \rangle$ correlation can be detected with high significance, it may be more
appropriate to split the estimator in Eq.~(\ref{eq:ttg_estimator}) into $\ell$ bands and estimate the cross
power spectrum $C_\ell^{\phi g}$ in independent bandpowers.\footnote{Let us mention one more subtlety
in the $TTg$ estimator (Eq.~(\ref{eq:ttg_estimator})):
inverse noise weighting the ${\hat\phi}$ field is only optimal if the lens reconstruction is
noise-dominated ($C_\ell^{\phi\phi} \ll N^{\phi\phi}_{\ell}$).  If this assumption is not satisfied
then the choice of optimal estimator will depend on whether optimality is defined assuming a fiducial
model with $C_\ell^{\phi\phi}=0$.  If $C_\ell^{\phi\phi}|_{\rm fid}$ is assumed zero (which makes
sense if one is trying to obtain the most statistically significant detection of a nonzero
$\langle TTg \rangle$ signal) then the estimator in Eq.~(\ref{eq:ttg_estimator}) is optimal.  If
$C_\ell^{\phi\phi}|_{\rm fid}$ is assumed nonzero (which makes sense if one is trying to obtain
the smallest error bars on $C_\ell^{\phi g}$ bandpowers around their fiducial values), then
the inverse noise weighted field $({\mathcal N}^{-1} {\hat\phi})$ should be replaced by a (suitably defined)
inverse signal+noise weighted field $({\mathcal S} + {\mathcal N})^{-1} {\hat\phi}$ in Eq.~(\ref{eq:ttg_estimator}).
A similar comment applies to the four-point estimator in Eq.~(\ref{eq:estimator_tttt}).
This subtlety will be unimportant for the WMAP analysis in the next subsection where the lens
reconstruction is very noise-dominated.}

Finally, we consider the four-point CMB signal given by Eq.~(\ref{eq:lensing_tttt}).
A nearly-optimal estimator is the auto power spectrum of ${\hat\phi}$, summed over $\ell$ in one
bandpower with ``shape'' proportional to the signal power spectrum $C_\ell^{\phi\phi}$:
\begin{equation}
\widehat{\mathcal E} = \frac{1}{F} \sum_{\ell m} C_\ell^{\phi\phi} 
  \left[ ({\mathcal N}^{-1} {\hat\phi})_{\ell m} ({\mathcal N}^{-1} {\hat\phi})_{\ell m}^*  
    - \left\langle ({\mathcal N}^{-1} {\hat\phi})_{\ell m} ({\mathcal N}^{-1} {\hat\phi})_{\ell m}^* \right\rangle
  \right]
\label{eq:estimator_tttt}
\end{equation}
where $\langle\cdot\rangle$ denotes an expectation value taken over random realizations of the CMB.
As with the $TTg$ estimator, we can alternately split the $\ell$ sum into subranges and estimate the
power spectrum $C_\ell^{\phi\phi}$ in independent bandpowers.
Note that this estimator $\widehat{\mathcal E}$ is written so that it looks like a two-point quantity,
but is actually a four-point estimator in the CMB.
This estimator is ``nearly optimal'' because the optimal estimator for $C_\ell^{\phi\phi}$ contains
additional two-point counterterms which reduce the variance in the same way that including the one-point
counterterm in the $f_{NL}$ estimator (Eq.~(\ref{eq:fnlhat})) reduces the variance, relative to an estimator defined
with only a three-point term.  For more details see \citep{Smith:2010}.

In summary, one can construct higher-point estimators for CMB lensing in two mathematically equivalent ways.
Formally, one can write down expressions for three-point or four-point correlation functions generated by lensing,
and obtain optimal estimators as special cases of general expressions for the optimal estimators.
More intuitively, one can apply the quadratic lens reconstruction estimator $\widehat \phi$ (Eq.~(\ref{eq:phihat}))
and then compute cross and auto power spectra.
In lens reconstruction language, the three-point and four-point correlations generated by lensing can be
interpreted as either the two-point correlation between $\widehat \phi$ and another field, or the two-point 
function of $\widehat \phi$ itself.

\subsection{Gravitational lensing in WMAP}
\label{ssec:wmap_lensing}

In this subsection we will present lensing results from 3-year WMAP data, originally reported in \citet{Smith:2007rg}.
At WMAP resolution, the lens reconstruction ${\hat\phi}$ is highly noise-dominated.  Forecasting shows that there is
insufficient signal-to-noise to detect lensing using the auto power spectrum $C_\ell^{{\hat\phi}{\hat\phi}}$
(i.e. the estimator defined in Eq.~(\ref{eq:estimator_tttt})).
However, the signal-to-noise can be boosted by cross correlating ${\hat\phi}$ with another field which is highly
correlated to $\phi$ and less noisy.
The best candidate for such a cross-correlation is the radio galaxy number density from the NVSS survey \citep{Condon:1998}.
The NVSS survey has large sky coverage ($f_{\rm sky}=0.8$), low Poisson noise ($N_{\rm gal} = 1.8 \times 10^6$), and
high redshift ($z_{\rm median} = 0.9$), making it an excellent match to a large-$f_{\rm sky}$ CMB lens reconstruction.

The main computational problem in implementing the optimal $TTg$ estimator
is the $C^{-1}$ operation appearing in Eq.~(\ref{eq:ttg_estimator}).
However, a solution to this problem has already been described (in the context
of primordial non-Gaussianity) in \S\ref{ssec:fnl_wmap}

Another implementational issue is that the definition of ${\hat\phi}$ given above
would be computationally prohibitive (the computational cost would be ${\mathcal O}(\ell_{\rm max}^5)$)
if evaluated in harmonic space using Eq.~(\ref{eq:phihat}).
However, a fast mathematically equivalent expression for ${\hat\phi}$ is given by the following chain of definitions:
\begin{eqnarray}
\alpha({\bf n}) &=& \sum_{\ell_1m_1} (C^{-1}a)^{\ell_1m_1} Y_{\ell_1m_1}({\bf n}) \nonumber \\
\beta_i({\bf n}) &=& \sum_{\ell_2m_2} C_{\ell_2}^{TT} (C^{-1}a)^{\ell_2m_2} \nabla_i Y_{\ell_2m_2}({\bf n}) \nonumber \\
({\mathcal N}^{-1} {\hat\phi})_{\ell m} &=& \int d^2{\bf n}\, (\nabla^i Y_{\ell m}^*({\bf n})) \Big[ \alpha({\bf n}) \beta_i({\bf n}) - \langle \alpha({\bf n}) \beta_i({\bf n}) \rangle \Big] 
\end{eqnarray}
This is closely analogous to the $f_{NL}$ estimator considered previously: the most algebraically
straightforward way to write down the estimator (Eq.~(\ref{eq:fnlhat})) has computational cost ${\mathcal O}(\ell_{\rm max}^5)$, 
but there is a faster algorithm based on the specific form of the three-point function.

\begin{figure}[!ht]
\begin{center}
\plotone{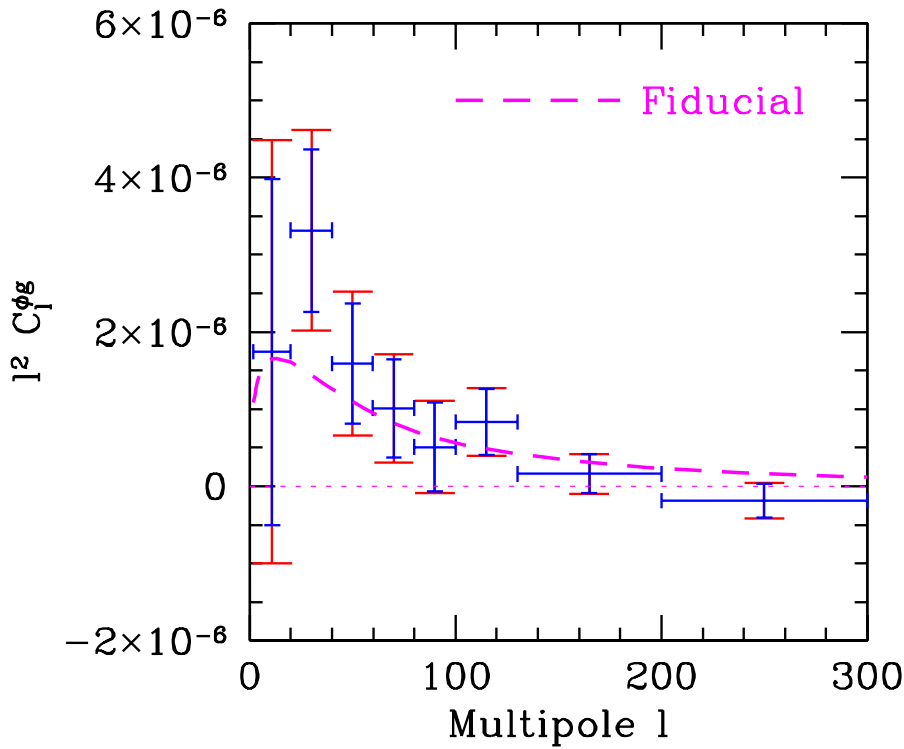}
\caption{Detection of gravitational lensing in WMAP, obtained by cross-correlating the lens reconstruction ${\hat\phi}$ with
galaxy number counts from NVSS using the optimal estimator (Eq.~(\ref{eq:ttg_estimator})).
The estimated power spectrum $C_\ell^{\phi g}$ is consistent with the fiducial model in the WMAP3 cosmology.
The two sets of error bars represent statistical and (statistical + systematic) errors.
Sources of systematic error include
number density gradients in NVSS, beam effects in WMAP, Galactic microwave foregrounds, resolved and unresolved CMB point sources, 
and the thermal Sunyaev-Zeldovich effect.
The significance (statistical+systematic) of the detection is 3.4$\sigma$, by fitting the estimated bandpowers
to an overall multiple of the fiducial power spectrum $C_\ell^{\phi g}$.
(From \citet{Smith:2007rg}.)}
\end{center}
\end{figure}

In Fig.~1, we show the result of evaluating the optimal $(TTg)$ estimator (Eq.~(\ref{eq:ttg_estimator}))
using CMB measurements from WMAP and galaxy number counts from NVSS.
We have split the estimator into several $\ell$ bands and reported an estimate for $C_\ell^{\phi g}$ in each band.
The errorbars include systematic errors from number density gradients in NVSS, beam effects in WMAP, CMB foregrounds
and point sources, and thermal SZ.
The overall significance (statistical+systematic) for detecting nonzero $C_\ell^{\phi g}$ is 3.4$\sigma$.
This result was the first detection of CMB lensing.  A detection was also reported by \citet{Hirata:2008cb}, using
a suboptimal estimator but a larger galaxy sample obtained by combining NVSS with luminous red galaxies and photometric
quasars from SDSS.

\section{Discussion and future prospects}

The theme of this article has been higher-point CMB signals which complement the power spectrum.
We have studied examples of such signals from inflationary physics and gravitational lensing in the late 
universe, constructed optimized estimators which are ``matched'' to various higher-point signals, and
reported results from WMAP data.
Let us conclude with a look to the future, by discussing upcoming observational prospects and some of the many
unsolved theoretical problems in this area.

For single-field inflation, there is a theorem (Eq.~(\ref{eq:F_single_field})) which completely characterizes
the most general three-point function which can be generated during inflaton.
This has not yet been generalized to multifield inflation.  The local shape
$f_{NL}^{\rm local}$ is one example of a three-point signal which can arise in multifield models but is disallowed in the
single-field case, but are there other possibilities?
Is it always sufficient to look for primordial three-point signals, or do there exist inflationary models which generate
detectably large four-point signals with no accompanying detectable three-point signal?


So far we have not mentioned CMB polarization.
Future generations of low-noise polarization experiments will be exquisite probes of CMB secondaries such as lensing,
since first-order perturbative effects with scalar sources only generate an E-mode in polarization 
\citep{Seljak:1996gy,Kamionkowski:1997},
whereas secondary effects tend to generate a mixture of E and B-modes.
For example, in the limit of low instrumental noise, the reconstructed lens potential ${\hat\phi}$ from CMB polarization
extends to much smaller angular scales than would be possible using CMB temperature \citep{Okamoto:2003zw}.
As another example, secondary polarization generated by Thomson scattering of CMB photons by HII bubbles during
the epoch of inhomogeneous reionization is a mixture of E and B-modes and generates a higher-point signal which
can be extracted by suitably constructed estimators \citep{Dvorkin:2008tf}.

At the time of this writing, CMB lensing has been detected in cross-correlation with large-scale structure,
with low statistical significance (3.4$\sigma$).
The situation will change dramatically in a few years with lensing results from Planck and ground-based experiments
like SPT and ACT, which can constrain lensing
``internally'' (i.e. without a cross-correlation tracer) at the few percent level and obtain interesting constraints
on parameters such as neutrino mass.
The CMB lens reconstruction ${\hat\phi}$ from Planck will be the first all-sky lensing map with interesting signal-to-noise,
and will measure a redshift range which is difficult to measure with other probes of lensing such as cosmic shear.
Gravitational lensing will soon be an indirect but important scientific product of experiments which measure the small-scale 
CMB, in much the same way that weak lensing (via galaxy ellipticities) is an important product of 
wide-field optical surveys like SDSS.

On small angular scales, secondary CMB anisotropy generated well after recombination dominates the primary CMB.
In this regime, the CMB should be thought of in a different way: each source of secondary anisotropy is a non-Gaussian signal
whose statistical distribution is different from the other secondaries.
Exploring this new observational frontier will require new statistical tools; the higher-point estimators
presented here represent one approach to this problem.
Ideally, we would like to have a complete set of estimators which
can separate the various secondaries (lensing, inhomogeneous reionization, kinetic SZ, etc.) from each other,
but such a framework has not yet been developed.
There are many unsolved problems in the field, but the potential scientific returns from new measurements of CMB
temperature and polarization in the next few years are very exciting.

\end{document}